\documentclass[reprint,aps,prxquantum,nofootinbib,superscriptaddress]{revtex4-2}

\usepackage{graphicx}
\usepackage{amsmath,amssymb}
\usepackage{xcolor}
\usepackage{hyperref}
\hypersetup{linkbordercolor={1 1 1}, urlbordercolor={1 1 1}}
\usepackage{booktabs}

\graphicspath{{figures/}}

\newcommand{\gross}{[[144,12,12]]}
\newcommand{\twogross}{[[288,12,18]]}
\newcommand{\bbninety}{[[90,8,10]]}
\newcommand{\hgp}{[[450,32,8]]}

\begin{document}

\title{BP-based qLDPC Decoding under Long-range Correlated Noise: Trapping Configurations and Prior Mismatch}

\author{Abhinav Dangi}
\affiliation{Maharshi Dayanand University, Rohtak, Haryana 124001, India}
\author{Avhishek Chatterjee}
\affiliation{Indian Institute of Technology Madras, Chennai 600036, India}

\begin{abstract}

We study the robustness of two state-of-the-art BP-based qLDPC decoders, Relay-BP and beam search, designed for independent noise when applied to spatially correlated noise with unchanged single-mechanism error probabilities. We generate correlated errors with a hidden Gaussian Markov random field, which lets us control the strength and range of the correlation while keeping every marginal error probability exact. This setting complements correlation-aware decoding: it asks how existing decoders perform when their assumed joint error model is mismatched. Across multiple tested circuits and code instances that we simulated, we observe that correlations can change logical error rates by orders of magnitude, reduce the finite-size distance advantage, and alter the relative ranking of Relay-BP, beam search, and BP-OSD. On one of the two tested circuits, Relay-BP and beam search become worse than BP-OSD, and increasing relay memory or beam width does not compensate for the degradation. Configuration analysis indicates that correlation primarily increases the probability of trapping configurations that are known to  challenge these decoders; at matched error weight, correlated errors still cause decoding failures several times more often than i.i.d.\ errors. Finally, supplying the decoders with the marginals conditioned on the hidden field recovers most of the lost performance in the tested settings identifying prior mismatch as a major contributor to performance degradation under correlation.

\end{abstract}

\maketitle

\section{Introduction}
\label{sec:intro}

Quantum low-density parity-check (qLDPC) codes can offer fault-tolerance with lower space overhead than geometrically local codes \cite{bravyi2024gross,tillich2014hgp,panteleev2021osd}, but the success of their practical deployment depends on fast and accurate decoding. Belief propagation (BP) has low computational cost and compatibility, but its performance on quantum LDPC codes is limited by degeneracy, short cycles, and associated trapping configurations\cite{raveendran2021trapping}. Ordered-statistics post-processing (BP-OSD) \cite{panteleev2021osd,roffe2020decoding} provides higher accuracy. However, its high computational time motivated explorations for faster alternatives.

Two recent BP-based approaches are seen as promising. Relay-BP \cite{muller2025relay} uses a sequence of BP instances with disordered memory strengths to perturb the BP dynamics and escape configurations in which ordinary BP becomes trapped \cite{muller2025relay}. On the other hand, beam-search decoding \cite{ye2025beam} identifies unreliable  variables(e.g., oscillating ones) from BP trajectories and branches on selected (most unreliable) variables, keeping a fixed number of competing hypotheses (the beam width). Both approaches improve substantially over conventional BP under the independent circuit-level noise models used in their original evaluations.

Relay and  beam search 
(a.k.a. branch-and-fix) mechanisms are  tuned and validated entirely on
independent noise. In practice, for quantum hardware like superconducting, neutral atom and photonics, correlated errors arise from spatially extended interactions, crosstalk, leakage, environmental fluctuations, shared control mechanisms, and other device-level effects. Hence, whether these decoders that make real-time implementations of qLDPC codes possible
 survive long-range correlated noise is a crucial question in quantum fault tolerance, which has received comparatively less attention.

Long-range correlated noise has also been studied analytically in quantum fault tolerance using pairwise joint Hamiltonian  models \cite{klesse2005correlated,aharonov2006correlated,preskill2013correlated} for geometrically local codes, and recently through hidden Markov random field models \cite{bagewadi2024correlated,kashtikar2026expander} that can approximate any stationary error distribution arbitrarily well \cite{KunschGK1995}. However, the decoders studied in these works are computationally complex or have lower noise thresholds. 

 Explorations into efficient quantum decoders against correlated errors have grown  recently. Correlated matching incorporates correlations between detection events in the MWPM matching graph while ensuring fast decoding \cite{PalerFowler2023CorrelatedMWPM}. More recently, search-based approaches like  Tesseract \cite{Beni2025Tesseract} and decision-tree decoding \cite{Ott2025DecisionTree} demonstrated good error suppression by explicitly exploring error configurations. Graph-based message-passing approaches have also been adapted to correlated errors. GARI rewired the decoding graph to remove problematic short cycles and combined it with normalized min-sum and ensemble decoding \cite{Maan2026GARI}. In another line of work, rather than modifying the decoding algorithm, the decoder is supplied with priors or model parameters that better represent the physical scenario \cite{Hockings2025NoiseAware}.

In practice, the effective noise environment of a quantum processor may not be
static and may change with operating conditions. It has been observed that temporal
drift and context dependence may arise from changes in control fields, 
environmental conditions, and other hardware-specific effects
\cite{Proctor2020Drift,Rudinger2019Context}. Correlated error mechanisms can also
arise from environmental fluctuations or radiation-induced events that change dynamically \cite{Wilen2021Correlated}. Recent work on quantum error correction has explicitly considered the problem of
decoding under drifting noise, highlighting the limitations of conventional
approaches based on stationary noise models and the potential benefits of adaptive
noise estimation \cite{Bhardwaj2026Adaptive}.
However, for  a practical fast decoder, accurately identifying and
continuously tracking the relevant correlation structure is substantially more
demanding than estimating the individual error probabilities used by the decoder.
 It is therefore important to
understand how well-established, computationally lightweight decoders perform when
the correlations are not known to the decoder.

\subsection*{Our Contribution}
We address this by studying relay-BP and beam search decoders in the presence of long-range correlated errors modeled using hidden Markov random field model \cite{bagewadi2024correlated,kashtikar2026expander}. Our error mechanism retains its original marginal probability exactly, while a hidden Gaussian Markov field induces correlations among errors at different qubit locations. Strength and range of correlation can be varied independently of the marginal error rate.

We evaluate the published configurations of relay-BP and beam search decoders without any alterations on correlated channels, and compare them with BP-OSD. The intention is to evaluate the robustness of the published decoders calibrated for i.i.d. channel.

First, we quantify the degradation produced by correlations at fixed marginal error rates across several qLDPC and surface-code circuits. Second, we investigate whether the degradation is simply due to an increase in the error weight by comparing correlated and i.i.d. error configurations. Third, we identify certain error configurations that cause  failures and test whether increasing the search budget allows the decoder to overcome them. Finally, we consider the case where the decoder is supplied with the marginals conditioned on the hidden field. This is motivated by the fact that correlated errors can be seen as heterogeneous noise whose marginals are sampled according to a hidden random field.

 We observe that correlation can lead to error configurations that are difficult for these BP decoders, causing logical error rates to increase by orders of magnitude even when the marginals are the same. Increasing search budget improves logical error rate under correlation, but its effect is far from compensating for the degradation due to correlation. Correlation strongly amplifies the probability of seeing decoder-failing configurations, i.e., error configurations from which the decoders cannot escape. These configurations remain fatal even when they are planted in decoders under uncorrelated noise.

 Our experiments also show a result complementary to the findings about error configurations. Much of the degradation of decoders' performance under correlation can be recovered when the marginals conditioned on the hidden field are supplied to the decoder as priors. Thus, these decoders are not intrinsically unsuitable for correlated noise, but  they can be sensitive to mismatch between static homogeneous marginals and the actual marginals conditioned on the hidden field. In other words, correlation makes these error configurations frequent but they defeat the iterative decoders mainly because the decoders lack knowledge of the actual marginals conditioned on the hidden field.

\subsubsection*{Organization of the paper}
In Sec.~\ref{sec:literature}, we discuss other lines of work that study effect of correlations on quantum fault-tolerance and position our work relative to them. Section~\ref{sec:model} introduces the correlated noise model, Sec.~\ref{sec:setup} the numerical setup and Sec.~\ref{sec:results} the performance of the decoders under correlated noise. Sections~\ref{sec:weight}--\ref{sec:oracle} address the mechanism: a matched-weight control (Sec.~\ref{sec:weight}), an analysis of decoder-failing configurations (Sec.~\ref{sec:traps}) and decoding with marginals conditioned on the hidden field (Sec.~\ref{sec:oracle}). Section~\ref{sec:discussion} concludes.

\section{Related Literature}
\label{sec:literature}
There are a few lines of research that explicitly incorporate correlations into
quantum error decoding. However, the current paper explores a different direction. Here, we discuss the existing work and position our work in that context.

A recent BP decoder, GARI, modifies the BP graph based on the knowledge of correlation with the goal of 
eliminating certain short cycles  \cite{Maan2026GARI}. It combines the transformed graph with
normalized min-sum decoding and randomized ensembles. 
In \cite{Beni2025Tesseract}, the decoding problem is presented as an A$^*$ search over subsets of errors that are consistent with the syndrome and can handle multiple correlated errors introduced by a common event., i.e., a single fault that flips several detectors, rather than correlations among different error mechanisms as studied here. Another class of search based decoders that has been shown to correct certain correlated errors is the decision-tree  decoders in \cite{Ott2025DecisionTree}.

The well known minimum-weight perfect matching (MWPM) algorithm was extended by Paler and Fowler \cite{PalerFowler2023CorrelatedMWPM} to incorporate correlations between errors while retaining the speed of MWPM. Recently,  decoder weights were calibrated using the noise characterization obtained from averaged
circuit eigenvalue sampling \cite{Hockings2025NoiseAware}. Learning-based decoders have 
demonstrated the benefits of adapting to device-specific noise. For example, AlphaQubit runs pretraining on simulated data followed
by fine-tuning on real data 
\cite{GoogleQuantumAI2024AlphaQubit}.

Existing work on correlated errors attempts to calibrate decoders to exploit correlation structures. However, in practice, the correlation structure may change with operating conditions. This motivates the current work where the robustness of known decoders against correlation is studied using controlled experiments that lead to possible remedies.

\section{Correlated noise model}
\label{sec:model}

We work at the level of a circuit's detector error model (DEM): a set of error
mechanisms $j = 1, \dots, M$, each with marginal probability $p_j$, a set of
detectors it flips, and a set of logical observables it flips. Under the
i.i.d.\ assumption of \cite{muller2025relay,ye2025beam}, errors fire
independently. We replace this with a concrete instantiation of the
generalized hidden-MRF model from \cite{bagewadi2024correlated,kashtikar2026expander}. 

It is known that hidden MRF models can approximate any stationary and ergodic distribution arbitrarily well \cite{KunschGK1995}. We use a hidden Gaussian Markov field with binary observables to model correlated errors. This model provides a tractable way to control the strength of the spatial  correlation while retaining the exact marginal probabilities.

\emph{Hidden field.} For each syndrome-extraction round $t$, an independent
stationary Gaussian AR(1) chain over the qubit index $k$:
\begin{equation}
J_{t,1} \sim \mathcal{N}(0,1), \qquad
J_{t,k+1} = \rho\, J_{t,k} + \sqrt{1-\rho^2}\, \xi_{t,k},
\end{equation}
with $\xi_{t,k} \sim \mathcal{N}(0,1)$ i.i.d., correlation length
$L = n^{\gamma}$ over the $n$ qubits, and $\rho = e^{-1/L}$. The exponent
$\gamma \in [0,1]$ mirrors the total-correlation scaling $V(J) \sim n^{\gamma}$
of Refs.~\cite{bagewadi2024correlated,kashtikar2026expander}: $\gamma \to 0$ is
short-range, $\gamma = 1$ is system-wide correlation, and $\gamma = 1/2$
(correlation length $\sqrt{n}$) is the critical scaling identified in
~\cite{kashtikar2026expander}.

\emph{Error firing (Gaussian copula).} Each DEM mechanism $j$ is mapped to a
representative qubit $q_j$ and round $t_j$ using Stim's error-mechanism
explanations. Given the field $J$, mechanisms fire \emph{conditionally
independently} with probability
\begin{equation}
g_j(J) \;=\; \Phi\!\left(
\frac{\Phi^{-1}(p_j) + \theta\, J_{t_j, q_j}}{\sqrt{1-\theta^2}}
\right),
\label{eq:copula}
\end{equation}
where $\Phi$ is the standard normal CDF and $\theta \in [0,1)$ is the copula
loading. This link has three properties that make the comparison clean.

First, $\mathbb{E}_J[g_j(J)] = p_j$ for all $\theta$: every
 marginal probability is preserved exactly, with no clipping bias.
Thus, the comparison isolates the effect of changing the joint distribution while holding all marginals fixed. The correlated and i.i.d. channels therefore have identical error rates but generally different error-weight distributions and higher-order spatial statistics.
Second, errors are conditionally independent given $J$, which is exactly the
hidden-MRF structure $q(Y \mid X) = \prod_i q_i(Y_i \mid X)$ of
Ref.~\cite{bagewadi2024correlated}, with $g_j$ smooth in $J$ as the theory
requires. Third, $\theta = 0$ reduces to i.i.d.\ sampling exactly, so the
published benchmarks of both decoder papers sit at the origin of our parameter
plane $(\theta, \gamma)$.

Rounds use independent hidden chains in this work.  Temporal correlation between
rounds is an interesting topic for future work. The hidden chain runs over the qubit
\emph{index}, not a geometric layout; for the bivariate-bicycle and
hypergraph-product codes studied here there is no planar embedding to prefer,
and we regard the ordering as part of the model definition
(Sec.~\ref{sec:discussion}).

The sampler is validated by a test suite checking exact marginal preservation,
the prescribed covariance decay, determinism under seeding, and bit-exact
agreement with i.i.d.\ sampling at $\theta = 0$; end-to-end, the $\theta = 0$
pipeline reproduces the stock samplers of both decoder repositories within
statistics (Sec.~\ref{sec:setup}).

In this paper, indexing of the hidden field variables is defined along the qubit ordering used by the circuit representation rather than along a hardware-specific geometric layout. This allows the correlation range to be controlled independently of the code geometry and makes the construction applicable to arbitrary DEMs. It is not a model for any particular hardware platform and offers a generic way of looking at correlation. Therefore, the results we obtain should be interpreted  as a test of decoder robustness to structured correlations rather than as a prediction for a particular hardware platform. 
To test whether  ranking of decoder performance depends on this embedding, we  repeat selected experiments after randomly permuting the qubit-to-field assignment. The resulting changes in absolute logical error rates do not alter the qualitative conclusions about the loss in performance of the decoders.

\section{Numerical setup}
\label{sec:setup}

\emph{Codes and circuits:} Throughout, $[[n,k,d]]$ denotes a code on $n$
physical qubits encoding $k$ logical qubits with distance $d$. We use the
exact memory-experiment circuits
distributed with each decoder paper's repository: from
Ref.~\cite{muller2025relay}, the \gross{} (``gross'') and \twogross{}
(``two-gross'') bivariate-bicycle codes \cite{bravyi2024gross} and a $d=11$
rotated surface code \cite{dennis2002topological,kitaev2003ftanyons}; from
Ref.~\cite{ye2025beam}, the \gross{} and \bbninety{} bivariate-bicycle codes
and a \hgp{} hypergraph-product code \cite{tillich2014hgp}. Each circuit runs
$d$ rounds of syndrome extraction at circuit-level depolarizing noise of
strength $p$. X- and Z-basis memory experiments are decoded separately and
combined as $\mathrm{LER} = \mathrm{LER}_X + \mathrm{LER}_Z$ per round, the
convention of both papers.

\emph{Decoders:} Relay-BP in its published \texttt{relay-bp-1} configuration
(301 legs, ensemble size 1) \cite{muller2025relay}; the beam-search decoder at
beam width 8 with 230 iterations (\texttt{beam8}), width 32 with 340
(\texttt{beam32}), and width 64 with 640 (\texttt{beam64}); and the beam
paper's BP-OSD reference, 30 BP iterations with order-10 OSD
(\texttt{bp30+osd}) \cite{ye2025beam}. All configurations, priors, and code
paths are as published: in particular, every decoder is initialized with the
\emph{i.i.d.\ marginals} $p_j$, matching what a decoder calibrated on averaged
error rates would use in practice.

\emph{Protocol:} For each (code, decoder, basis, $p$, $\gamma$, $\theta$) we
sample syndromes from the hidden-MRF sampler and decode until 20\,000 shots or
300 logical failures, whichever comes first (relative standard error
$\lesssim 7\%$). Every decision-critical cell (the entries of
Tables~\ref{tab:gross} and \ref{tab:desat} and the $\gamma = 0.5$
$\theta$-sequence of Sec.~\ref{sec:ranking}) is then extended with a second,
independent seed to at least $1000$ and typically $2000$--$3100$ combined
failures ($2$--$3\%$ relative error), and the i.i.d.\ baselines to $10^5$
shots per basis; failure counts are quoted with each table. Since each shot
draws a fresh hidden field, shots are i.i.d.\ samples of the correlated
process and seeds simply extend the sample; the two seeds agree within
statistics at every cell checked.
Baselines reproduce the papers: our i.i.d.\ Relay-BP baseline on the gross
code at $p = 0.003$ gives $6$ failures in $10^4$ shots against the
repository's reference of $13$ in $10^4$ (consistent at these statistics), and
our i.i.d.\ beam-decoder baseline gives $\mathrm{LER} = 3.2 \times 10^{-4}$
per round against ${\sim}3 \times 10^{-4}$ in Fig.~2 of
Ref.~\cite{ye2025beam}. Passing $\theta = 0$ through the correlated pipeline
matches both stock pipelines ($7$ vs.\ $6$ failures per $10^4$ for relay;
$18$ vs.\ an expected $20.6$ per $10^4$ for beam).

\section{Performance under correlated noise}
\label{sec:results}

\begin{table*}
\caption{Logical error rate per round (X+Z) for the \gross{} code at
$p = 0.003$, i.i.d.\ ($\theta = 0$) versus strong correlation
($\theta = 0.9$) at three correlation ranges $\gamma$. Parenthesized factors
are relative to \texttt{bp30+osd} in the same column (values $< 1$: better
than BP-OSD). Two independent seeds per entry; $147$--$3141$ combined
failures per entry ($2$--$8\%$ relative error, $2$--$3\%$ on all correlated
cells). Published-pipeline comparison: each decoder runs on its own paper's
circuit; see Table~\ref{tab:singlecirc} for the single-circuit control.}
\label{tab:gross}
\begin{ruledtabular}
\begin{tabular}{lcccc}
Decoder & $\theta=0$ & $\theta=0.9$, $\gamma=0.25$ & $\theta=0.9$, $\gamma=0.5$ & $\theta=0.9$, $\gamma=1.0$ \\
\hline
Relay-BP (\texttt{relay-bp-1}) & $1.2\times10^{-4}$ (0.22) & $3.3\times10^{-2}$ (3.8) & $2.9\times10^{-2}$ (0.59) & $1.3\times10^{-2}$ (0.21) \\
Beam search (\texttt{beam8})   & $3.2\times10^{-4}$ (0.58) & $9.8\times10^{-3}$ (1.12) & $4.9\times10^{-2}$ (1.00)  & $6.0\times10^{-2}$ (0.96) \\
BP-OSD (\texttt{bp30+osd})     & $5.5\times10^{-4}$ (1)    & $8.7\times10^{-3}$ (1)   & $4.9\times10^{-2}$ (1)    & $6.2\times10^{-2}$ (1) \\
\end{tabular}
\end{ruledtabular}
\end{table*}

\begin{table*}
\caption{Companion to Table~\ref{tab:gross} away from saturation: the same
comparison at $p = 0.001$, $\theta = 0.9$, where per-shot failure rates are
$4$--$29\%$ and the decoders have room to differ. Two seeds;
$218$--$2283$ failures per entry (the $\gamma = 0.25$ beam8 and BP-OSD
entries reached the $4\times10^4$-shot cap at ${\sim}220$ failures,
${\sim}7\%$ relative error; all others $2$--$3\%$). Published-pipeline
comparison, as in Table~\ref{tab:gross}.}
\label{tab:desat}
\begin{ruledtabular}
\begin{tabular}{lccc}
Decoder & $\gamma=0.25$ & $\gamma=0.5$ & $\gamma=1.0$ \\
\hline
Relay-BP   & $7.5\times10^{-3}$ (21.6) & $9.2\times10^{-3}$ (1.01) & $4.8\times10^{-3}$ (0.19) \\
Beam search & $4.5\times10^{-4}$ (1.31) & $8.4\times10^{-3}$ (0.93) & $2.4\times10^{-2}$ (0.96) \\
BP-OSD     & $3.5\times10^{-4}$ (1)    & $9.0\times10^{-3}$ (1)    & $2.5\times10^{-2}$ (1) \\
\end{tabular}
\end{ruledtabular}
\end{table*}

\subsection{Dependence on correlation strength}
\label{sec:strength}

\begin{figure*}
\includegraphics[width=\textwidth]{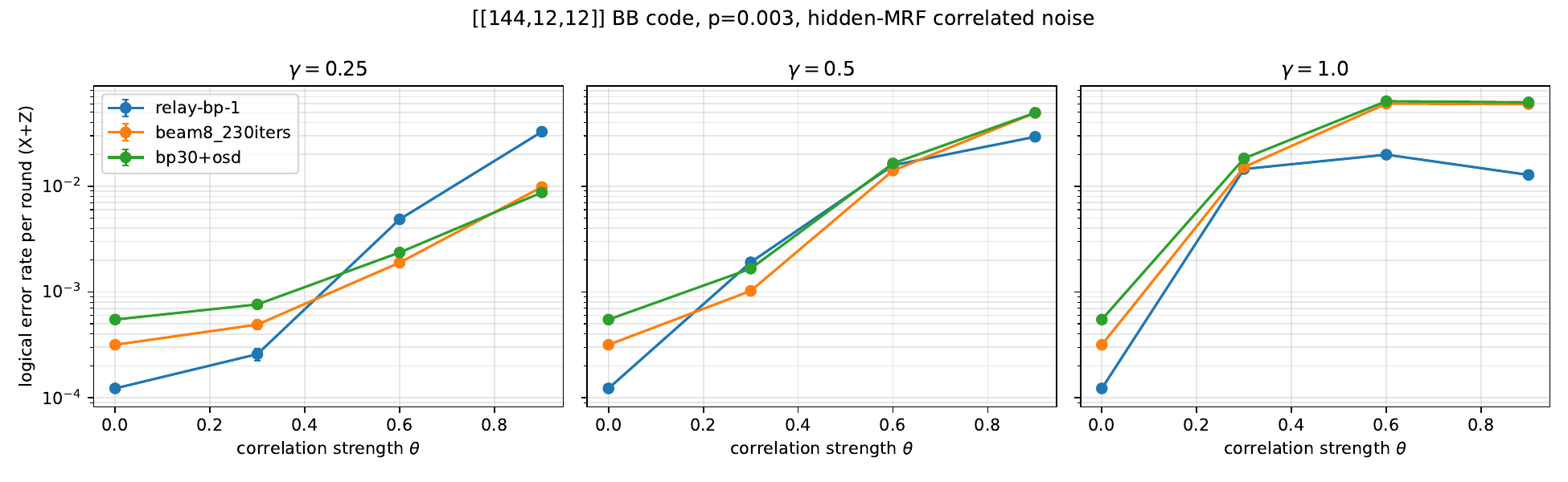}
\caption{Logical error rate per round (X+Z) versus correlation strength
$\theta$ for the \gross{} code at $p=0.003$, at three correlation ranges
$\gamma$. All marginal error probabilities are held exactly fixed. Error
bars: binomial $1\sigma$.}
\label{fig:grid}
\end{figure*}

Figure~\ref{fig:grid} and Table~\ref{tab:gross} show the central phenomenon on
the \gross{} code at $p = 0.003$. Every error mechanism's marginal
probability is identical in every column; only the correlation changes. At
$\theta = 0.9$, $\gamma = 0.5$ the logical error rate rises by a factor of
$240$ for Relay-BP, $160$ for beam search, and $90$ for BP-OSD. The effect
strengthens at lower physical error rate, where the i.i.d.\ code is deeper
below threshold: at $p = 0.001$ Relay-BP's i.i.d.\ rate is below the
resolution of $2\times10^4$ shots (no failures observed; $95\%$ upper bound
$2.5\times10^{-5}$ per round) while the correlated rate at $\theta = 0.6$ is
$1.2\times10^{-3}$ and at $\theta = 0.9$ is $9.2\times10^{-3}$, at least a
factor of 50 and 370, respectively, and consistent with three orders of
magnitude.

\subsection{Collapse of sub-threshold error suppression}
\label{sec:psweep}

\begin{figure}
\includegraphics[width=\columnwidth]{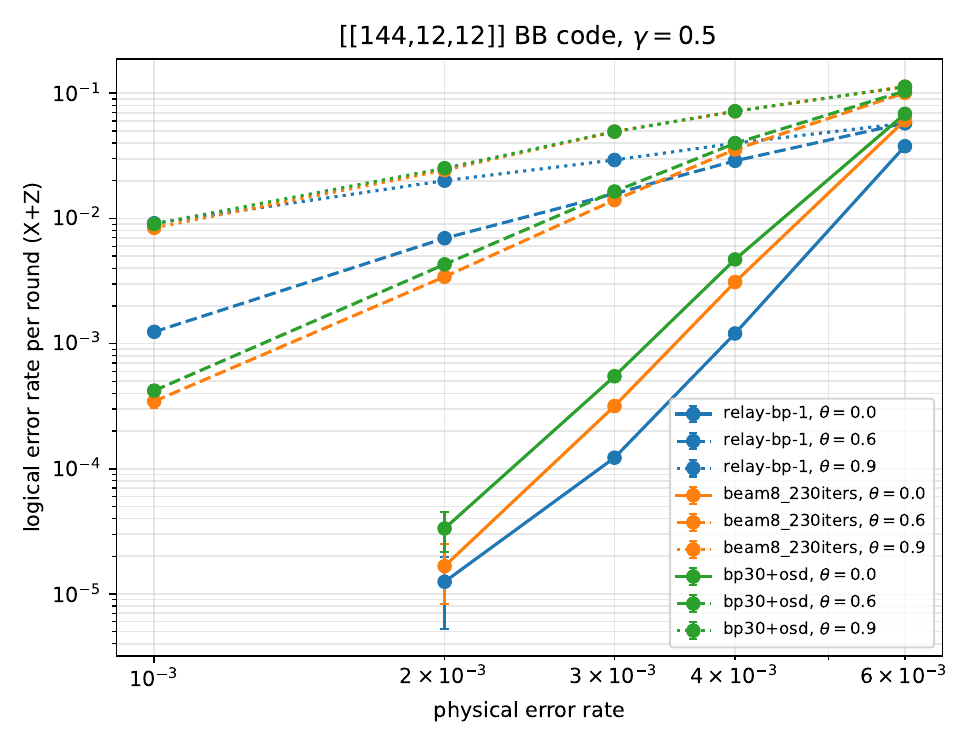}
\caption{Logical error rate per round versus physical error rate $p$ for the
\gross{} code at $\gamma = 0.5$ and $\theta \in \{0, 0.6, 0.9\}$. The steep
i.i.d.\ slope flattens as correlation grows. Error bars: binomial
$1\sigma$.}
\label{fig:psweep}
\end{figure}

Figure~\ref{fig:psweep} sweeps the physical error rate at $\gamma = 0.5$.
Under i.i.d.\ noise all three decoders show the steep sub-threshold
suppression that distance provides: Relay-BP's LER falls by more than three
orders of magnitude from $p = 0.006$ to $p = 0.001$ (from $3.8\times10^{-2}$
to below $2.5\times10^{-5}$). At $\theta = 0.9$ the same sweep spans about
a factor of six ($5.8\times10^{-2}$ to $9.2\times10^{-3}$). The
suppression exponent, the practical meaning of code distance, is largely lost under strong long-range correlation over the tested range of $p$, consistent with failures dominated by rare system-wide bursts rather than independent low-weight events.

\subsection{Decay and inversion of the decoder advantage}
\label{sec:ranking}

The parenthesized ratios in Tables~\ref{tab:gross} and \ref{tab:desat} carry
the message of this paper. Under i.i.d.\ noise the published rankings
reproduce: Relay-BP beats BP-OSD by $4.5\times$ and beam search by
$1.7\times$ on the gross code. Along the strength axis at $\gamma = 0.5$ the
two decoders lose their advantage in different ways. Beam search decays
monotonically to parity: its ratio to BP-OSD moves
$0.58 \to 0.62 \to 0.85 \to 1.00$ at $\theta = 0, 0.3, 0.6, 0.9$. Relay-BP
crosses below parity almost immediately: $0.22 \to 1.15 \to 0.96 \to
0.59$. Already at the weak correlation $\theta = 0.3$ it is $15\%$
\emph{worse} than plain BP-OSD ($1.9\times10^{-3}$ vs.\ $1.7\times10^{-3}$,
${\sim}4$ standard deviations), and its apparent recovery at $\theta = 0.9$
reflects BP-OSD saturating rather than relay improving. At strong
correlation all three decoders converge to within a factor of ${\sim}2$:
where correlation is strong, the choice of decoder has little effect.

More strikingly, the ranking inverts with the correlation range. Under
strong \emph{short-range} correlation ($\gamma = 0.25$, $\theta = 0.9$)
Relay-BP is the \emph{worst} of the three decoders, at $3.3\times10^{-2}$
against $8.7\times10^{-3}$ for BP-OSD, a factor of 3.8 \emph{behind} the
decoder it beats by $4.5\times$ under i.i.d.\ noise. Beam search is also
statistically behind plain BP-OSD there (ratio $1.12 \pm 0.03$). At
system-wide correlation ($\gamma = 1.0$) the order flips again: Relay-BP
($1.3\times10^{-2}$) is five times better than beam search and BP-OSD
(${\sim}6\times10^{-2}$), which saturate. The comparison away from saturation at
$p = 0.001$ (Table~\ref{tab:desat}) shows the inversion is not a saturation
artifact; saturation was in fact compressing it: away from saturation, Relay-BP is
$21.6\times$ worse than BP-OSD at $\gamma = 0.25$ and $5.2\times$ better at
$\gamma = 1.0$, while beam search tracks BP-OSD within $31\%$ everywhere.
Relay-BP is also non-monotonic in $\theta$ at $\gamma = 1$
($1.45\times10^{-2} \to 1.99\times10^{-2} \to 1.28\times10^{-2}$ at
$\theta = 0.3, 0.6, 0.9$, each entry at ${\sim}2\%$): system-wide bursts
polarize shots into nearly clean and catastrophic, and the relay mechanism
handles the clean majority well. Decoder robustness is correlation-regime
dependent; i.i.d.\ benchmark rankings do not transfer.

An important caveat governs the cross-decoder ratios above: following both
papers, each decoder runs on \emph{its own paper's} circuit (Relay-BP on the
circuit distributed with Ref.~\cite{muller2025relay}; beam search and BP-OSD
on the circuit distributed with Ref.~\cite{ye2025beam}). The two \gross{}
circuits implement different circuit-level noise recipes at the same nominal
$p$, and, because our hidden field chains over the qubit index, different
correlation embeddings. Tables~\ref{tab:gross} and \ref{tab:desat} therefore
compare \emph{published pipelines}, not decoding algorithms in isolation.
Section~\ref{sec:singlecirc} performs the controlled comparison.

\subsection{A single-circuit control for the ranking inversion}
\label{sec:singlecirc}

We re-ran the comparison with all three decoders decoding \emph{identical
shots} of a single circuit (Relay-BP accepts any detector error model, and
symmetrically the beam decoder and BP-OSD run on the relay paper's circuit),
on both papers' \gross{} circuits, with the hidden field additionally
re-embedded under random qubit-index permutations as a control
(Table~\ref{tab:singlecirc}). Three facts emerge.

\begin{table*}
\caption{Single-circuit control: per-shot failure probability (Z basis,
$p = 0.003$) when all three decoders decode identical shots of one circuit.
Parenthesized: ratio to BP-OSD. Each row $200$--$1100$ failures per decoder.
Permutation rows re-embed the hidden chain under a random qubit permutation
(two permutations shown pooled; individual permutations agree within
statistics).}
\label{tab:singlecirc}
\begin{ruledtabular}
\begin{tabular}{llccc}
Circuit & Setting & Relay-BP & Beam search & BP-OSD \\
\hline
beam  & i.i.d.                  & $7.7\times10^{-4}$ (0.22) & $2.2\times10^{-3}$ (0.63) & $3.5\times10^{-3}$ \\
beam  & $\theta{=}0.3$, $\gamma{=}0.5$  & $2.2\times10^{-3}$ (0.22) & $6.4\times10^{-3}$ (0.64) & $1.0\times10^{-2}$ \\
beam  & $\theta{=}0.9$, $\gamma{=}0.25$ & $2.3\times10^{-2}$ (0.50) & $4.5\times10^{-2}$ (1.00) & $4.5\times10^{-2}$ \\
beam  & \quad + permuted        & $8.3\times10^{-3}$ (0.35) & $2.0\times10^{-2}$ (0.87) & $2.3\times10^{-2}$ \\
beam  & $\theta{=}0.9$, $\gamma{=}1.0$   & $3.4\times10^{-1}$ (0.94) & $3.4\times10^{-1}$ (0.96) & $3.6\times10^{-1}$ \\
relay & i.i.d.                  & $7.8\times10^{-4}$ (0.22) & $2.5\times10^{-3}$ (0.73) & $3.5\times10^{-3}$ \\
relay & $\theta{=}0.3$, $\gamma{=}0.5$  & $1.2\times10^{-2}$ (0.44) & $2.4\times10^{-2}$ (0.88) & $2.7\times10^{-2}$ \\
relay & $\theta{=}0.9$, $\gamma{=}0.25$ & $1.7\times10^{-1}$ (2.51) & $2.5\times10^{-1}$ (3.67) & $6.9\times10^{-2}$ \\
relay & \quad + permuted        & $1.9\times10^{-1}$ (2.67) & $2.7\times10^{-1}$ (3.75) & $7.3\times10^{-2}$ \\
\end{tabular}
\end{ruledtabular}
\end{table*}

First, on a fixed circuit the escape mechanisms' advantage decays with
correlation but need not invert: on the beam paper's circuit, Relay-BP
remains the best decoder at every setting tested, its advantage over BP-OSD
shrinking from $4.5\times$ (i.i.d.\ and $\theta = 0.3$) toward parity at
strong correlation. The short-range inversion of Table~\ref{tab:gross} does
not reproduce there.

Second, the inversion is real, and on the circuit where it occurs it
affects both escape mechanisms: on the relay paper's circuit at
$\theta = 0.9$, $\gamma = 0.25$, plain BP-OSD is the best decoder by a wide
margin: Relay-BP fails $2.5\times$ and beam search $3.7\times$ more often
on the same shots. The branch-and-fix mechanism inverts \emph{harder} than
the relay mechanism. Whatever structure this circuit's error mechanisms
acquire under short-range correlation, BP-based escape strategies are
defeated by it while OSD post-processing is not. At weak correlation
($\theta = 0.3$) the same circuit shows no inversion, so the effect requires
strong, short-range correlation \emph{and} this circuit structure.

Third, the effect is not an artifact of how the hidden chain is
embedded: re-embedding the field under random qubit permutations changes
absolute failure rates (by ${\sim}2\times$ on the beam circuit, whose natural
ordering is measurably adversarial for correlated noise) but never the
within-circuit ranking, on either circuit. The inversion is a property of
the circuit's error-mechanism structure, invariant under re-embedding.

The practical statement for decoder benchmarking is sharper than a simple
ranking inversion: whether an i.i.d.-tuned escape mechanism survives
correlated noise depends on the circuit it decodes, and neither published
pipeline's behavior predicts the other's. A decoder qualified on one
circuit family under i.i.d.\ noise carries no guarantee under correlation
even relative to the simplest baseline.

\subsection{Correlation reduces the finite-size distance advantage}
\label{sec:distance}

\begin{figure}
\includegraphics[width=\columnwidth]{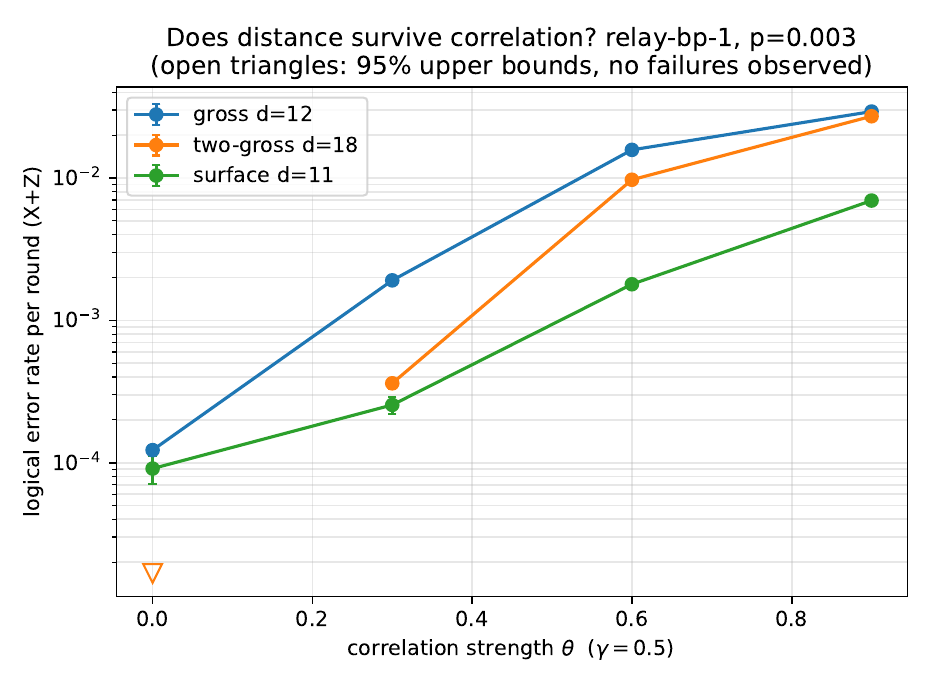}
\caption{Distance test: \gross{} ($d=12$) versus \twogross{} ($d=18$) under
Relay-BP at $p = 0.003$, $\gamma = 0.5$. The i.i.d.\ distance advantage
(leftmost points) collapses as $\theta$ grows. Open triangles: $95\%$
upper bounds where no failures were observed. Error bars: binomial
$1\sigma$.}
\label{fig:distance}
\end{figure}

Figure~\ref{fig:distance} compares the \gross{} ($d = 12$) and \twogross{}
($d = 18$) codes under Relay-BP at $\gamma = 0.5$, the critical
$\sqrt{n}$-range scaling of Ref.~\cite{kashtikar2026expander}. Under
i.i.d.\ noise the two-gross code shows no failures in $2\times10^4$ shots
where the gross code sits at $1.2\times10^{-4}$: an advantage of at least an
order of magnitude, consistent with the papers. The advantage shrinks to
$5.4\times$ at $\theta = 0.3$, $1.6\times$ at $\theta = 0.6$, and $1.1\times$
at $\theta = 0.9$ ($2.7\times10^{-2}$ vs.\ $2.9\times10^{-2}$), statistically
indistinguishable from no advantage at all. Doubling the qubit count to
reach $d = 18$ gives no measurable advantage at $\theta = 0.9$ in this comparison, consistent at finite size and at decoder level with the storage-exponent degradation proved for expander codes in Ref.~\cite{kashtikar2026expander}.

\subsection{Increased beam width does not recover the loss}
\label{sec:beamwidth}

If correlated failures were ordinary trapping sets of the kind the branching
mechanism targets, a wider beam should break more of them. It does not.
On the \gross{} code at $\gamma = 0.5$, widening the beam from 8 to 64 (and
raising the iteration budget from 230 to 640) improves the i.i.d.\ LER by a
factor of 3.3 ($3.2\times10^{-4} \to 9.6\times10^{-5}$), lifting the beam
decoder's advantage over BP-OSD from $1.7\times$ to $5.7\times$. At
$\theta = 0.9$ the same $8\times$ increase in search improves the LER by
$19\%$ ($4.9\times10^{-2} \to 4.0\times10^{-2}$), an advantage over BP-OSD of
$1.2\times$. The same pattern holds on the \bbninety{} and \hgp{} codes.
Whatever the structure of the correlated failures, additional search of the
kind the decoder performs barely affects them, a first indication that these
failures are not branch-recoverable (Sec.~\ref{sec:traps}).

\subsection{Results across all five codes}
\label{sec:allcodes}

Table~\ref{tab:allcodes} extends the comparison to the remaining codes, each
decoded by its own paper's decoder(s) on its own paper's circuits. The degradation is present in each of the five code instances tested, with instructive variation.
The \hgp{} hypergraph-product code ($d = 8$, $k = 32$) is hit hardest,
reaching $1.7\times10^{-1}$ per round at $\theta = 0.9$, LER per shot
${\sim}0.8$, effectively undecodable. This corner cannot be brought away
from saturation within the paper's methodology: at $p = 0.002$, the lowest error rate for
which Ref.~\cite{ye2025beam} publishes a \hgp{} circuit, the $\theta = 0.9$
per-shot failure rate is still ${\sim}1.0$; we therefore quote no decoder
ratios there. At $\theta = 0.6$, $p = 0.002$ the code does come off the
floor (per-shot rate $0.29$), and the beam decoder's edge over BP-OSD
($1.14\times$, $2\%$ statistics) persists. The
$d = 11$ surface code is the most robust in absolute terms
($6.9\times10^{-3}$ at $\theta = 0.9$), but it encodes $k = 1$ logical qubit
against $k = 8$--$32$ for the others, and normalizing per logical qubit
(Table~\ref{tab:perk}) reverses that impression: under i.i.d.\ noise the
qLDPC codes are $5$--$10\times$ better per logical qubit than the surface
code (the qLDPC advantage reported in the papers), and even at
$\theta = 0.9$ the bivariate-bicycle codes remain ${\sim}3\times$ ahead of
it, with all five codes compressed into a factor of ${\sim}3$ per logical
qubit. For the two tested bivariate-bicycle codes, the substantial logical-error advantage of increasing \(d\) under i.i.d. noise does not exist when there is a strong correlation. However, this does not prove that distance ceases to provide asymptotic protection under correlated noise. Establishing such a statement would require a more systematic study involving distance scaling.

\begin{table}
\caption{Per-logical-qubit logical error rate per round
($\mathrm{LER}/\mathrm{round}/k$) at $p = 0.003$, $\gamma = 0.5$, each code
on its own paper's circuit and decoder (relay-bp-1 for the first three,
beam8 for the last two). The two-gross $\theta = 0$ entry is an upper
bound.}
\label{tab:perk}
\begin{ruledtabular}
\begin{tabular}{lcccc}
Code & $k$ & $\theta=0$ & $\theta=0.3$ & $\theta=0.9$ \\
\hline
\gross{} BB      & 12 & $1.0\times10^{-5}$ & $1.6\times10^{-4}$ & $2.4\times10^{-3}$ \\
\twogross{} BB   & 12 & $<1.4\times10^{-6}$ & $3.0\times10^{-5}$ & $2.3\times10^{-3}$ \\
$d{=}11$ surface & 1  & $9.1\times10^{-5}$ & $2.5\times10^{-4}$ & $6.9\times10^{-3}$ \\
\bbninety{} BB   & 8  & $1.4\times10^{-4}$ & $3.3\times10^{-4}$ & $5.4\times10^{-3}$ \\
\hgp{} HGP       & 32 & $6.1\times10^{-5}$ & $2.4\times10^{-4}$ & $5.3\times10^{-3}$ \\
\end{tabular}
\end{ruledtabular}
\end{table}
Notably, the beam paper's internal ordering (beam8 ahead of BP-OSD by
$1.5$--$2\times$) is stable at every $\theta$ on both of its codes, unlike
the Relay-BP/beam inversion on the gross code: the fragility ordering is
decoder-pair and code dependent, which any hardware-facing benchmark would
need to map.

\begin{table}
\caption{Logical error rate per round (X+Z) at $p = 0.003$, $\gamma = 0.5$,
for the remaining code--decoder pairs (each code on its own paper's circuit
and decoder). The two-gross $\theta = 0$ entry is a $95\%$ upper bound (no
failures in $2\times10^4$ shots per basis).}
\label{tab:allcodes}
\begin{ruledtabular}
\begin{tabular}{llccc}
Code & Decoder & $\theta{=}0$ & $\theta{=}0.3$ & $\theta{=}0.9$ \\
\hline
\twogross{}    & relay-bp-1 & $<\!1.7{\times}10^{-5}$ & $3.6{\times}10^{-4}$ & $2.7{\times}10^{-2}$ \\
$d{=}11$ surface & relay-bp-1 & $9.1{\times}10^{-5}$ & $2.5{\times}10^{-4}$ & $6.9{\times}10^{-3}$ \\
\bbninety{}    & beam8      & $1.1{\times}10^{-3}$ & $2.7{\times}10^{-3}$ & $4.3{\times}10^{-2}$ \\
\bbninety{}    & bp30+osd   & $1.9{\times}10^{-3}$ & $3.9{\times}10^{-3}$ & $4.7{\times}10^{-2}$ \\
\hgp{}         & beam8      & $2.0{\times}10^{-3}$ & $7.5{\times}10^{-3}$ & $1.7{\times}10^{-1}$ \\
\hgp{}         & bp30+osd   & $3.1{\times}10^{-3}$ & $1.1{\times}10^{-2}$ & $1.8{\times}10^{-1}$ \\
\end{tabular}
\end{ruledtabular}
\end{table}

\subsection{Runtime under correlated noise}
\label{sec:runtime}

\begin{figure*}
\includegraphics[width=\textwidth]{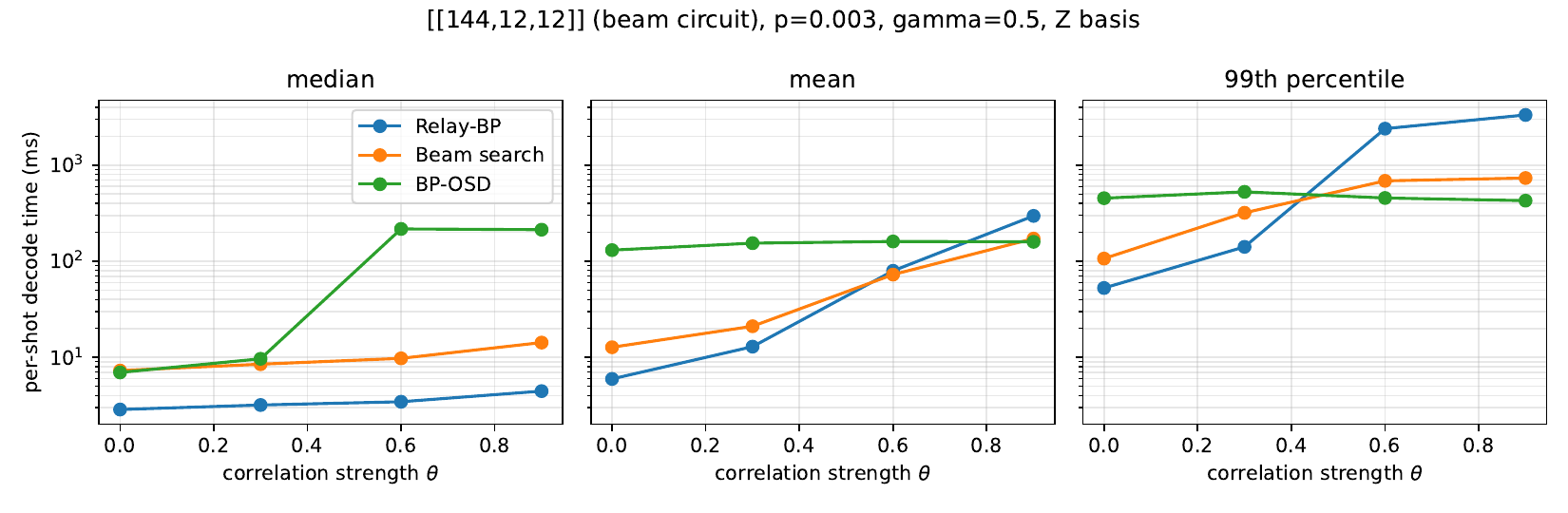}
\caption{Per-shot decoding time versus correlation strength $\theta$
(\gross{} beam circuit, $p = 0.003$, $\gamma = 0.5$, Z basis, $2000$ shots
per point, single-threaded latency; measured concurrently on 8 worker
processes, so ratios rather than absolute times are meaningful). Median,
mean, and $99$th percentile. Relay-BP's median remains flat while its mean
and tail grow by factors of $50$--$60$: under strong correlation a shot is
typically either trivial or worst-case.}
\label{fig:runtime}
\end{figure*}

Both papers argue for real-time viability from the tail of the per-shot
decoding-time distribution under i.i.d.\ noise, where the typical shot
converges in a few BP iterations and expensive escapes are rare.
Figure~\ref{fig:runtime} measures that distribution under correlation
(single-threaded per-shot latency on the \gross{} beam circuit at
$\gamma = 0.5$; absolute values are platform-specific, ratios are the
point). Under i.i.d.\ noise the escape mechanisms pay only in the tail:
Relay-BP's median is $2.9$~ms with a $99$th percentile of $53$~ms, and
BP-OSD's median $7$~ms hides a $130$~ms \emph{mean} driven by rare OSD
invocations. At $\theta = 0.9$ this balance inverts. Relay-BP's mean
per-shot time grows $50\times$ (to $297$~ms) and its $99$th percentile
$63\times$ (to $3.3$~s), while its median stays at $4$~ms: shots
polarize into clean shots that decode immediately and worst-case shots that
traverse all 301 relay legs, mirroring the burst polarization of
Sec.~\ref{sec:ranking}.
BP-OSD's median rises $30\times$ (to $213$~ms): the typical shot now
invokes OSD. Beam search's mean grows $14\times$. A decoder budgeted for
the i.i.d.\ tail falls behind the syndrome stream precisely when the noise
is correlated; any real-time claim needs a correlated-tail budget.

\section{A matched-weight control: error shape versus error weight}
\label{sec:weight}

A skeptical reading of Sec.~\ref{sec:results} attributes everything to error
weight: correlation at fixed marginals overdisperses the number of physical
errors per shot, occasionally exceeding what \emph{any} decoder could correct,
so the LER collapse would be information-theoretic rather than algorithmic.
Because the sampler knows the injected error $e$ on every shot, this is
directly testable: we bin shots by injected error weight and compare the
conditional failure probability $P(\mathrm{fail} \mid \mathrm{weight})$ under
correlated versus i.i.d.\ sampling.

\begin{figure*}
\includegraphics[width=\textwidth]{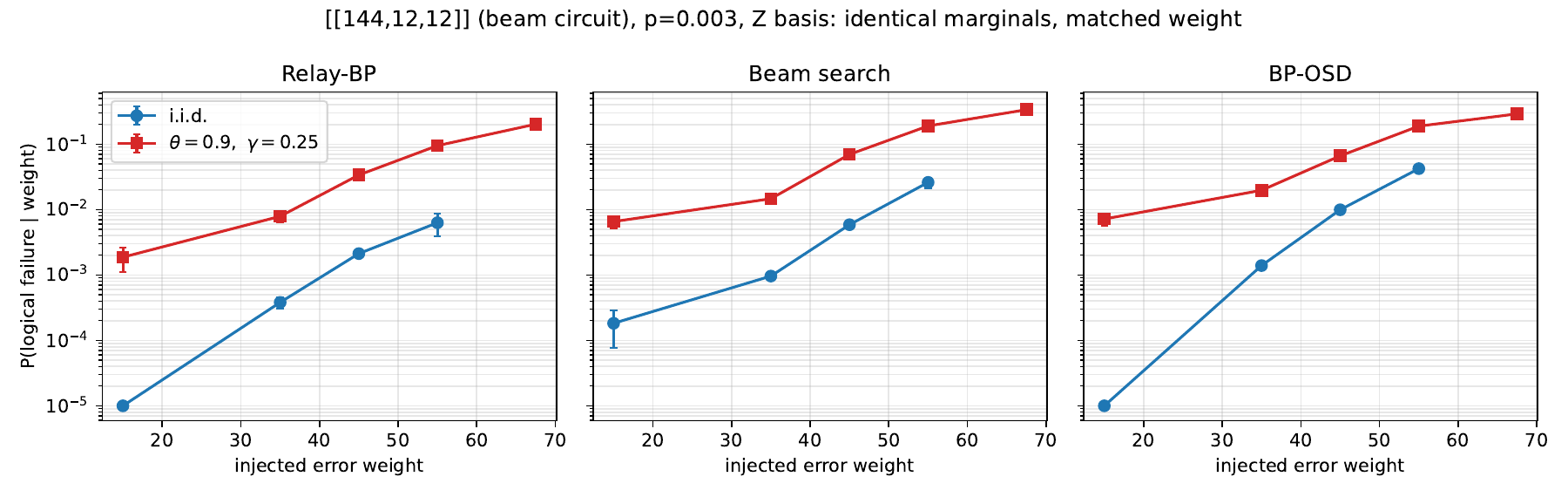}
\caption{Matched-weight control on the \gross{} (beam) circuit at
$p = 0.003$ (Z basis): conditional failure probability versus injected error
weight, i.i.d.\ versus strongly correlated sampling at identical marginals
(mean injected weight $35.3$ in both). At matched weight, clustered
errors defeat all three decoders $4$--$36\times$ more often than scattered
errors of the same weight, in every weight bin where both ensembles have failures.}
\label{fig:weight}
\end{figure*}

The extra weight of correlated errors does not account for the degradation. On the \gross{} circuit at $p = 0.003$ (Z basis,
$10^5$ i.i.d.\ shots vs.\ $10^4$ shots at $\theta = 0.9$, $\gamma = 0.25$),
the mean injected weight is $35.3$ in both ensembles, the exact marginal
preservation of Eq.~\eqref{eq:copula} at work. Yet at \emph{matched} weight
the correlated shots fail an order of magnitude more often: for weights
below 45, which carry the bulk of both distributions, the beam decoder's
conditional failure probability is $1.5\times10^{-3}$ under i.i.d.\ sampling
and $1.8\times10^{-2}$ under correlation ($12\times$); in the 45--55 bin,
$1.1\times10^{-2}$ vs.\ $1.2\times10^{-1}$ ($11\times$). The correlated
ensemble does also have a fatter weight tail (weights above 65 essentially
never occur under i.i.d.\ sampling), and those tail shots fail at
$50$--$100\%$; but the tail carries a small fraction of the correlated
failures. A typical-weight error, re-arranged into clusters by the hidden
field, is what the decoders cannot handle, the object Sec.~\ref{sec:traps}
characterizes. The same conclusion holds for BP-OSD and Relay-BP
(Fig.~\ref{fig:weight}); the conditional-failure gap is largest at \emph{low}
weight, where the decoders almost never fail on i.i.d.\ errors.

\section{Decoder-failing configurations under correlated noise}
\label{sec:traps}

Both decoders are belief propagation underneath, and differ only in the
escape strategy for trapped configurations
\cite{raveendran2021trapping,muller2025relay,ye2025beam}: Relay-BP perturbs
(disordered memory, fresh legs), beam search branches (clamp the least
reliable variable, mask, re-run). The beam paper's mechanism analysis
identifies one trap family (short-cycle LLR oscillations) and demonstrates,
under i.i.d.\ noise, that branching escapes it. Neither paper examines the nature of
trapping configurations that correlated noise throws at these decoders. This
section is devoted to studying them on the \gross{} code. We use the term trapping configuration for these objects as a regular phrase, rather than as the rigorous graph-theoretic definition in  \cite{raveendran2021trapping}.

\emph{Method:} Because the sampler knows the injected error on every shot,
failures can be analyzed without inference. (i) \emph{Harvest}: collect all
logical failures at $\theta = 0$ and at $(\theta, \gamma) \in
\{0.6, 0.9\} \times \{0.25, 0.5\}$, logging the injected error, syndrome, and
decoder telemetry. (ii) \emph{Shrink}: greedily remove mechanisms from each
failing configuration while the failure persists, yielding a minimal failing
configuration; deduplicate by Tanner-graph support. (iii) \emph{Label}: decode
each canonical configuration's exact syndrome with Relay-BP across its escape
budget (legs, ensemble, memory disorder) and with beam search at widths
8/32/64, labeling it \emph{relay-breakable} / \emph{branch-breakable} or not.
(iv) \emph{Tie to correlation}: measure each canonical family's occurrence
rate as a function of $(\theta, \gamma)$; and \emph{plant} minimal
correlated-noise configurations into the i.i.d.\ ($\theta = 0$) circuit to test whether they still cause failures without correlation.

All results below are single-circuit (Sec.~\ref{sec:singlecirc}): every
decoder decodes the same shots, Z basis, $p = 0.003$. We harvested $100$
failures per decoder per setting, shrunk each to a 1-minimal configuration
(batched greedy deletion; every reported configuration verifiably fails and
loses failure under any single-mechanism removal), deduplicated, and decoded
each canonical configuration's exact syndrome with the battery
$\{$beam8, beam32, beam64, BP-OSD, relay with $1/60/301$ legs$\}$. Because
Relay-BP draws random memory strengths per instantiation, its labels are
failure \emph{probabilities} over five independent instances: on marginal
configurations a fresh relay instance succeeds or fails stochastically, a
fact that any ``relay breaks it'' claim must respect.

\emph{The i.i.d.\ advantage is escapability:} Of minimal configurations that
defeat beam8 under i.i.d.\ noise, full Relay-BP escapes $96\%$ and beam64
escapes $96\%$; of those that defeat full Relay-BP, beam8 escapes $87\%$.
The escape mechanisms do exactly what their papers claim, on the trap
population that i.i.d.\ noise produces.

\emph{Correlation produces traps that no tested mechanism escapes:} For minimal
configurations harvested at $\theta = 0.9$, $\gamma = 0.25$ on the same
circuit, the unbreakable fractions rise threefold and more: Relay-BP fails
on $13\%$ of correlated beam-traps (vs.\ $4\%$ i.i.d.), beam64 on $15\%$
(vs.\ $4\%$); beam8 fails on $32\%$ of correlated relay-traps (vs.\ $13\%$).
The escape budgets also saturate: on this circuit relay with 60 legs and
301 legs perform identically on these configurations, and beam32 $\approx$
beam64, echoing Sec.~\ref{sec:beamwidth}: more of the same search does not
touch the correlated family. Weighting by the harvest rates, the rate of
\emph{jointly unbreakable} failures (fatal to both papers' mechanisms at
full budget) rises ${\sim}45\times$, from $1.1\times10^{-4}$ per shot at
$\theta = 0$ to $5.0\times10^{-3}$ at $\theta = 0.9$, and its share of all
failures roughly doubles.

\emph{Planting: the configurations fail without correlation:} We
injected the thirty hardest correlated-noise configurations into otherwise
i.i.d.\ ($\theta = 0$) noise, $200$ background samples each. Against an
ambient failure probability of ${\sim}2\times10^{-3}$, the planted shots
fail at mean probability $0.55$--$0.69$ per decoder on the beam circuit
($90\%$ of configurations remain majority-fatal for beam search and BP-OSD)
and up to $0.90$ on the relay circuit, a $10^2$--$10^{2.5}$-fold
amplification with no correlation present whatsoever. Since the planted configurations were chosen from correlated-noise failures, this shows that they suffice to cause failures, not that they are the only route by which correlation degrades decoding. It supports the interpretation that, at fixed marginals, the hidden field raises
the probability of clustered sub-configurations that are decoder-failing, rare under i.i.d.\ sampling, and not resolved by the tested escape mechanisms. Planted failure probabilities sit below one because background
noise occasionally rescues a shot: additional syndrome weight can move BP
out of the trapped fixed point.

\emph{The inversion circuit fails differently:} On the relay circuit at the
same setting (where both escape mechanisms fall behind plain BP-OSD,
Table~\ref{tab:singlecirc}), the traps are \emph{shallow but frequent}:
minimal configurations that defeat beam8 there are broken by beam64
($100\%$), Relay-BP ($100\%$), and even plain BP-OSD ($97\%$); yet beam8
fails a quarter of all shots, because the correlated field produces such
configurations at a very high rate. Each decoder has its own large,
mostly disjoint failure family (cross-decoder escape $60$--$70\%$), the
families are heavier objects than on the beam circuit (median minimal
weight $23$ vs.\ $11$), and unlike on the beam circuit, additional relay
legs \emph{do} help ($P(\mathrm{fail})$ drops $0.74 \to 0.36$ from 60 to 301
legs). The aggregate inversion is thus a \emph{frequency} phenomenon (which
trap families the correlated field presents to each decoder, and how often), not
an unbreakability phenomenon, consistent with OSD's robustness there: OSD's
post-processing does not depend on BP convergence, so it is insensitive to
the shallow-trap family that overwhelms the BP-based escape mechanisms.

\begin{figure*}
\includegraphics[width=\textwidth]{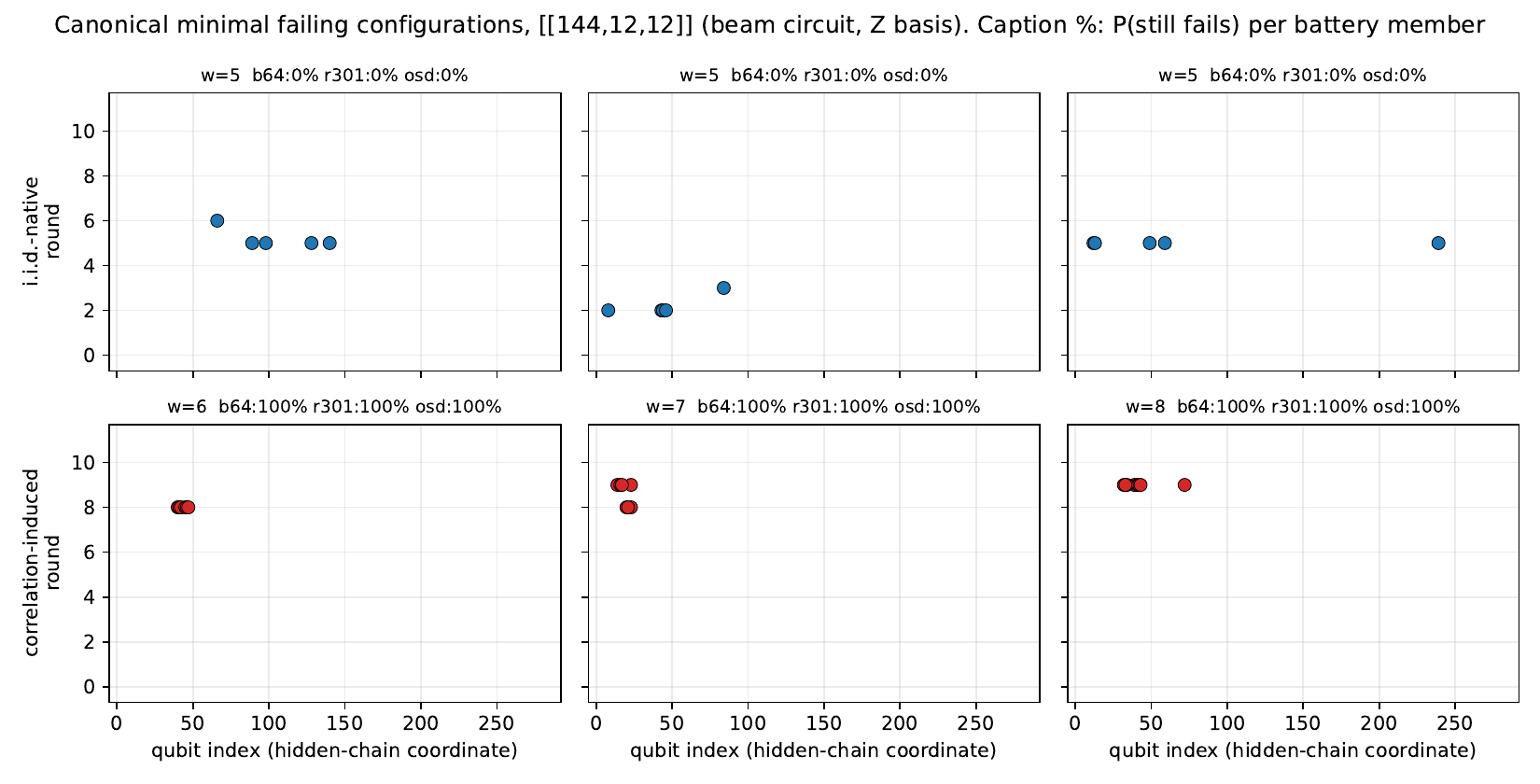}
\caption{Canonical minimal failing configurations on the \gross{} (beam)
circuit, Z basis, drawn in (qubit index, round) space. Top: i.i.d.-native
traps; scattered, low weight, broken by every escape mechanism
(percentages: probability the battery member still fails). Bottom: the
hardest correlation-induced traps; compact clusters of weight $6$--$8$
that defeat beam64, 301-leg Relay-BP, and BP-OSD with probability one.}
\label{fig:gallery}
\end{figure*}

\emph{What the traps look like:} Figure~\ref{fig:gallery} draws exemplar
minimal configurations in the hidden-chain coordinate. The contrast is the
mechanism made visible: i.i.d.-native traps are \emph{scattered} (weight-5
configurations spread across the code, and every escape mechanism breaks
them), while the hardest correlation-induced traps are \emph{compact
clusters} of weight only $6$--$8$, spanning a few tens of chain sites in a
single round, fatal to beam64, 301-leg relay, and OSD alike. The fatal
configurations are not large; they are dense.

\emph{A code-level hypothesis, tested and refuted:} The beam decoder
hard-codes a refusal to branch on variables of Tanner degree $\le 2$; one
might suspect correlated failures hide on exactly those variables. The
opposite holds: while trap support overall is mildly enriched on
degree-$\le 2$ columns ($14$--$19\%$ vs.\ a $9.8\%$ baseline), the
beam64-\emph{unbreakable} correlated traps are \emph{depleted} there
($4.2\% \pm 1.6\%$). The escape-proof configurations live on high-degree
variables that branching is fully permitted to act on, and fails anyway.
The failure is not the branching filter but the search landscape the
cluster creates.

\section{Oracle priors: decoding with the realized noise field}
\label{sec:oracle}

This section explores a direction complementary to the explorations so far. We observed in Sec.~\ref{sec:traps} that degradation under correlated noise is associated with amplification of probabilities of trapping configurations under high correlation. Here we ask the question: is it a fundamental phenomenon that all BP decoders in general, and relay-BP and beam search in particular, cannot avoid?

The hidden Gaussian field model can also be seen as follows. Independent and non-identically distributed binary errors are sampled from heterogeneous marginals $\{g_j(J)\}$, where the marginals are functions of a Gaussian field $J$. Essentially, a correlated error distribution is nothing but heterogeneous independent distributions, where the marginals change according to  the hidden field. Thus, if the hidden field is known, then the heterogeneous marginals are also known. If we assume that a hypothetical oracle knows $J$ and supplies $g_j(J)$ to the decoders, how does the performance change when decoders use these actual dynamic marginals (conditioned on $J$) instead of the static marginals $\mathbb{E}_J[g_j(J)]$?

The outcome is largely positive: in the tested settings, most of the
correlated-noise degradation is prior mismatch. Each shot
is decoded twice on identical syndromes (paired comparison; mechanisms map
one-to-one onto decoder columns for these circuits, and the oracle arm is
validated shot-exact against the static arm at $\theta = 0$).

At the strongest short-range point on the beam circuit ($\theta = 0.9$,
$\gamma = 0.25$; $5000$ shots), the static decoders fail on $2$--$4\%$ of
shots; with oracle priors, all three decoders achieve zero
failures: $0/5000$ each, better than their own i.i.d.\ performance
(${\sim}10$ expected failures in $5000$ shots), because the realized prior
tells the decoder where the burst sits. At $\gamma = 0.5$, where static
failure rates reach $22$--$29\%$ per shot, oracle priors reduce failures by
factors of $60$--$160$ (beam8 $833 \to 12$, BP-OSD $858 \to 15$, Relay-BP
$663 \to 4$ per $3000$), leaving residuals at i.i.d.\ levels. Even the
trapping sets of Sec.~\ref{sec:traps} lose their effect: the configurations
are fatal under the i.i.d.\ prior, not under the true one.

On the inversion circuit (relay circuit, $\theta = 0.9$, $\gamma = 0.25$)
the oracle removes most of the effect (Relay-BP $581 \to 18$, BP-OSD
$232 \to 1$ per $3000$) but not all of it: beam search retains $59/3000$
($2\%$) failures \emph{with exact priors}, Relay-BP retains $18/3000$, about $8\times$ its i.i.d.\ rate on this circuit, and the within-circuit ordering
of Table~\ref{tab:singlecirc} persists in miniature. The shallow-trap flood
that defeats branching on this circuit is therefore mostly, but not
entirely, a prior effect; a small circuit-specific weakness, largest for branching,
survives perfect side information.

Three qualifications are in order. The oracle knows the hidden field
$J$, not the error $e$: it is legitimate side information, but side
information no deployed decoder has; the result is a diagnostic upper bound
identifying \emph{field estimation} as the target, not a decoder. The
factorized form matters: BP exploits per-mechanism priors natively, so no
algorithmic change was needed; the entire gain comes through the channel
probabilities. And the recovery bound is tight only as far as tested: one
code family, space-only correlations, Z basis.

Thus, the decoders with i.i.d.-calibrated static marginals
fail under correlated noise, not the BP-based decoders as such. Thus, correlation determines how often the trapping configurations occur while the mismatch of marginals contributes significantly to the degradation in performance. %Given the marginals conditioned on $J$, the same configurations no longer defeat the decoders.

We would like to emphasize that this oracle experiment is a hypothetical experiment used to gain insight. We do not claim that the hidden field can be reliably estimated from the syndrome alone. An experimentally useful decoder that defeats correlation would have to estimate the  hidden field from side information like soft readout, leakage indicators, calibration data, or syndrome history. Designing such an estimator is outside the scope of this work.

\section{Discussion and Conclusion}
\label{sec:discussion}

As a complement to previous works on decoders for correlated noise, we study the robustness of decoders designed for i.i.d.\ noise under correlated noise. This approach is motivated by the possibility that in practice, correlations may change with time and the advantage  of the decoder may diminish if the correlation model used for calibration does not match the operating conditions; this work itself models spatial correlations only. We use a hidden Gaussian AR(1) field to model correlated errors because it allows the strength and range of the correlation to be controlled at exact marginals (Sec.~\ref{sec:model}).

Our experiments support three important conclusions. First, benchmarking
qLDPC decoders under independent noise measures their best case. Even for fixed
marginal error rates, correlation moves logical error rates by orders of
magnitude, substantially reduces the finite-size distance advantage in the tested comparison, and reorders the performance (LER) ranking of the BP
decoders.  Second, the advantage of the relay and
branch-and-fix (beam search) decoders over plain BP-OSD decays  as
correlation strength increases even if the marginals remain the same.  The experiments also show that on one of the two tested circuits Relay-BP and beam search can become worse than BP-OSD, under strong short-range correlation.

Third, deterioration in performance with increasing correlation cannot be compensated by increasing the search budget like beam-width or relay memory (Sec.~\ref{sec:beamwidth}).
These observations indicate that the escape mechanisms introduced by Relay-BP and beam search are sensitive to the error configurations encountered by the decoder.

Our configuration analysis provides a more specific interpretation. Strong correlation increases the frequency of error configurations that are difficult for these BP-based decoders. Many such configurations remain difficult when embedded into an otherwise i.i.d. background, indicating that their effect on the decoders does not require correlation to be present. The role of correlation is then mainly  to change how frequently these trapping configurations occur.

The oracle-prior experiment (marginals conditioned on the field)  gives a complementary conclusion. When the marginals conditioned on the  hidden field are supplied to the decoder, the LERs fall to or below the i.i.d. level on the beam circuit and by more than an order of magnitude on the relay circuit. This indicates that most of the observed degradation in the tested settings is due to prior mismatch: the decoder is calibrated to the unconditional marginals, while the actual error probabilities change with the latent field. The experiment shows how much of the performance can be recovered by providing better instantaneous noise information.

These observations position the current work alongside, and not in opposition to, recent correlation-aware decoding methods like correlated matching and GARI. Those approaches modify decoders to handle correlated errors, whereas our intention is to quantify the robustness of existing i.i.d.-calibrated BP-based decoders when the assumed joint distribution is wrong but all marginal error rates remain unchanged. This is motivated by the fact that the performance of the correlation-aware decoder modifications depends on how well the assumed correlation structure is tracked, and recalibration has a cost. However, in practice, since many temporal factors influence correlations in quantum systems, we investigate the complementary direction: the robustness of decoders against correlations of varying strength and range. 

A direction for future work is to replace the oracle with an estimator of the hidden field. Such an estimator could use soft measurement information, leakage or burst indicators, calibration streams, or temporal correlations in the syndrome record. The natural next question is whether a lightweight estimator of the latent field can provide a significant part of the oracle improvement. If the hidden field has strong temporal correlation in practice, an estimator that uses past syndrome data could provide a useful estimate of the conditional marginals; testing this requires extending our model, whose rounds are independent.

\begin{acknowledgments}
AC thanks ANRF India (formerly, SERB India) for support through grant CRG/2023/005345 and the Ministry of Education, India.
\end{acknowledgments}

\bibliography{references}

\end{document}